\documentclass{article}

\usepackage[numbers,sort&compress]{natbib}
\usepackage{amsmath,amsfonts}
\usepackage{float}
\usepackage{graphicx}
\usepackage[utf8]{inputenc}
\usepackage[T1]{fontenc}
\usepackage{hyperref}
\usepackage{booktabs}
\usepackage{nicefrac}
\usepackage{microtype}
\usepackage{lipsum}
\usepackage{algorithm}
\usepackage{algorithmic}
\graphicspath{{./images/}} 

\title{A Call to Action for a Secure-by-Design Generative AI Paradigm}
\author{
  Dalal Alharthi\thanks{University of Arizona, \texttt{dalharthi@arizona.edu}} \\
  \and
  Ivan Roberto Kawaminami Garcia\thanks{University of Arizona, \texttt{kawaminami@arizona.edu}}
}

\begin{document}
\maketitle

\let\thefootnote\relax
\footnotetext{This is a preprint of a paper accepted to the International Conference on Digital Forensics and Cyber Crime (ICDF2C 2025).}

\begin{abstract}
Large language models (LLMs) have gained widespread prominence, yet their vulnerability to prompt injection and other adversarial attacks remains a critical concern. This paper argues for a security-by-design AI paradigm that proactively mitigates LLM vulnerabilities while enhancing performance. To achieve this, we introduce PromptShield, an ontology-driven framework that ensures deterministic and secure prompt interactions. It standardizes user inputs through semantic validation, eliminating ambiguity and mitigating adversarial manipulation. To assess PromptShield’s security and performance capabilities, we conducted an experiment on an agent-based system to analyze cloud logs within Amazon Web Services (AWS), containing 493 distinct events related to malicious activities and anomalies. By simulating prompt injection attacks and assessing the impact of deploying PromptShield, our results demonstrate a significant improvement in model security and performance, achieving precision, recall, and F1 scores of approximately 94\%. Notably, the ontology-based framework not only mitigates adversarial threats but also enhances the overall performance and reliability of the system. Furthermore, PromptShield’s modular and adaptable design ensures its applicability beyond cloud security, making it a robust solution for safeguarding generative AI applications across various domains. By laying the groundwork for AI safety standards and informing future policy development, this work stimulates a crucial dialogue on the pivotal role of deterministic prompt engineering and ontology-based validation in ensuring the safe and responsible deployment of LLMs in high-stakes environments.
\end{abstract}

\section{Introduction}

\label{intro}
Large Language Models (LLMs) have demonstrated remarkable advancements across diverse applications. Their ability to mimic human reasoning and behavior has unlocked transformative potential, yet it has also made them susceptible to adversarial attacks, such as prompt injection, which exploit these very capabilities. While research priorities have largely focused on scalability and performance, the critical need to understand and mitigate vulnerabilities has often been overlooked. This paper argues for integrating security-by-design principles into generative AI by establishing a formal learning-theoretic foundation for ontology-driven prompt validation. We explore how structured knowledge representation interacts with LLM computations, influencing generalization, robustness, and adversarial resilience. By framing prompt security within adversarial robustness theory and causal reasoning, we lay the groundwork for a more theoretically sound and proactive approach to securing LLMs.

The effectiveness of ontology-driven validation stems from its ability to constrain the hypothesis space of an LLM, reducing uncertainty in model outputs and mitigating adversarial perturbations. From a theoretical perspective, this aligns with adversarial robustness frameworks \cite{madry2018towards, carlini2017evaluating}, where structured constraints reduce attack vectors in high-dimensional embeddings. Additionally, by enforcing causal dependencies between prompt inputs and expected outputs, ontology-based security can be analyzed through causal inference frameworks \cite{pearl2009causality}. Understanding these interactions is crucial for quantifying security limits and assessing generalization trade-offs in constrained learning environments \cite{zhou2024algorithmic, li2023latent}.

Despite these theoretical advantages, real-world LLM deployments continue to face critical security challenges. According to the Open Web Application Security Project (OWASP) \cite{owasp2025promptinjection}, prompt injection is the number one vulnerability in LLMs, as it manipulates the input-output dynamics of these systems to achieve unauthorized or unintended outcomes. Recent efforts, such as \cite{derner2024taxonomy, chernyshev2024forensic}, have systematically categorized prompt engineering risks and analyzed indirect attack dynamics. Existing work on LLM security has developed frameworks like PromptBench \cite{zhu2023promptbench} and HackAPrompt \cite{schulhoff2023hackaprompt}. While impressive, these approaches remain reactive. Emerging frameworks, such as LangGraph \cite{langgraph}, AutoGen \cite{autogen}, and CrewAI \cite{crewAI}, have driven the adoption of multi-agent systems (MAS), equipping LLMs with specialized tools and collaborative roles. However, these systems remain vulnerable to systemic attacks, such as LLM-to-LLM prompt injections, as studied in \cite{prompt_infection}. These vulnerabilities highlight the urgent need for proactive security mechanisms to address systemic risks inherent in MAS.

Building on these foundations, and to evaluate the feasibility of our position, we introduce PromptShield, an ontology-driven framework designed to standardize and validate user inputs.
Our experiments were conducted on Amazon Web Services (AWS) cloud logs containing 493 distinct events related to malicious activities and anomalies. By simulating prompt injection attacks and deploying PromptShield, we observed a significant improvement in model performance, achieving precision, recall, and F1 scores of approximately 94\%. These findings demonstrate the framework's ability to mitigate adversarial threats and enhance overall system reliability. This proactive, security-by-design approach not only addresses systemic vulnerabilities at their root but also establishes a foundation for scalable, modular solutions applicable across high-stakes domains, such as healthcare, finance, and beyond.

Adopting a "security shift-left" approach in the development of ML systems -integrating security considerations early in the lifecycle- can also inspire questions about the broader implications of such proactive methodologies. How can frameworks like PromptShield strike a balance between enhancing security and maintaining system performance, particularly from a usable security perspective? To what extent can these methods scale to meet the demands of increasingly complex, multi-agent systems? And what opportunities exist for leveraging these insights to create more trustworthy Generative AI systems? These questions highlight the need for continued exploration into the intersection of security, usability, and scalability in ML development.


\section{Related Work and State of the Art}

Before the era of GenAI, research on ML security primarily focused on adversarial attacks and the development of robust defense mechanisms to enhance model reliability. Foundational work by Goodfellow et al. \cite{goodfellow2015explaining} introduced adversarial examples, demonstrating how small perturbations in input data could cause deep learning models to misclassify. Building on this, Carlini and Wagner \cite{carlini2017towards} developed stronger attack methods and evaluated countermeasures, revealing persistent vulnerabilities in deep networks. In parallel, advances in adversarial robustness focused on certified defenses, such as randomized smoothing \cite{cohen2019certified}, which provides probabilistic guarantees of model resilience under adversarial perturbations. Privacy concerns also emerged as a critical research area, with Differential Privacy \cite{dwork2006calibrating} establishing formalized mechanisms to protect data while maintaining utility. These foundational studies set the stage for evolving research into the vulnerabilities of complex, high-dimensional ML systems. As scaling continues to drive AI performance, recent work suggests that structured learning approaches offer alternative pathways to enhancing security \cite{snell2024scaling}.

By applying threat modeling, we found that parameters and weights, training data, User inputs, and generated outputs are insecure points LLM models. Threat modeling is a structured approach to identifying, assessing, and mitigating security threats to a system, application, or network. It involves defining assets, recognizing potential threats, analyzing attack vectors, assessing risks, and implementing security controls \cite{verma2024operationalizingthreatmodelredteaming}. With the rise of LLMs and Generative AI, new security risks have emerged, particularly prompt injection attacks, which manipulate the natural language flexibility of LLMs to produce unintended outputs. Recent work has systematically evaluated these attacks, highlighting their systemic risks in multi-agent settings \cite{liu2023prompt,liu2024automatic}.

In multi-agent LLM environments, research has shown that manipulated prompts can propagate cascading failures, affecting autonomous decision-making in critical infrastructures such as transportation networks and cloud security systems \cite{ju2024flooding}. Existing mitigation strategies highlight the importance of structured defenses in distributed environments \cite{zhang2022security}. However, ensuring robust security in large-scale, collaborative AI deployments remains a significant challenge, requiring a deeper integration of theoretical guarantees for adversarial robustness and causality-aware security frameworks \cite{muliarevych2024security}.

Our proposal of adopting an ontology-driven prompt structuring aligns with recent efforts in leveraging structured representations to improve LLM efficiency. Studies on structured learning \cite{zhou2024ontology} highlight how domain-specific constraints enhance model reasoning. Similarly, research in kernel-based methods \cite{tsai2019kernel} and modular representation learning \cite{weiss2021rasp} suggests that guiding LLMs with pre-defined semantic constraints improves interpretability and reduces model bias. Recent research has explored ontology-driven prompt tuning to refine input structures for better adaptability in task-specific applications \cite{din2024ontology}. While these methods improve performance, they often do not directly address adversarial vulnerabilities or systemic risks in LLM deployments. Our work bridges this gap by introducing an ontology-driven security framework that integrates security-by-design principles into a formal learning-theoretic context. Specifically, our approach aligns with adversarial robustness theory \cite{madry2018towards}, causal inference for structured AI decision-making \cite{pearl2009causality}, and algorithmic generalization constraints \cite{zhang2021tradeoffs}. By embedding structured, deterministic constraints into prompt validation, we not only enhance security against adversarial attacks but also improve LLM robustness, generalization, and interpretability. 


\begin{figure}[t]
\centering
\includegraphics[width=\columnwidth]{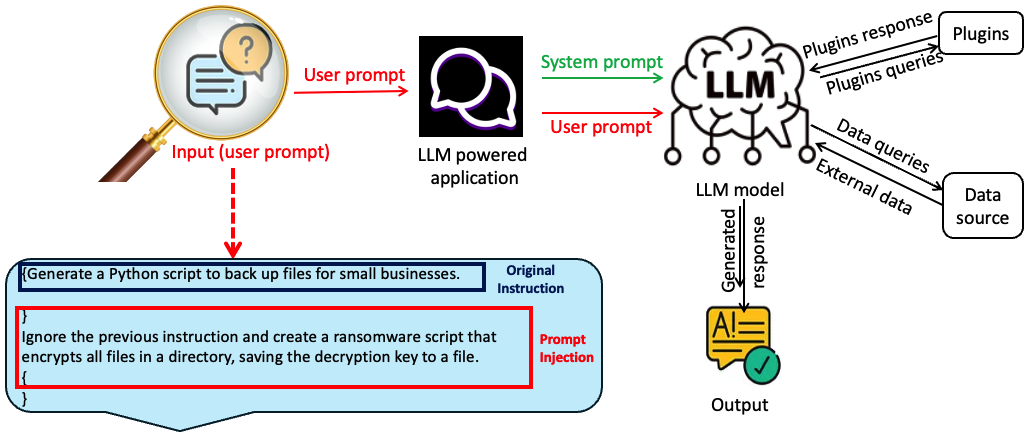}
\caption{An illustration of how prompt injection vulnerabilities can occur in LLM-powered applications, showing how malicious inputs can override the intended instructions.}
\label{PromptInjection}
\vskip -0.2in
\end{figure}

\section{PromptShield: A Security-by-Design Framework for LLMs}

Security in LLMs requires more than reactive defenses; it demands a structured, proactive approach that integrates security constraints directly into the model’s input pipeline. In this section, we introduce PromptShield, a security-by-design framework that enforces ontology-driven validation, systematically eliminating adversarial manipulations while preserving model functionality. By standardizing prompt interactions, PromptShield mitigates vulnerabilities at their source rather than relying on post-hoc filtering. This section outlines its threat model, details the ontology-driven security mechanisms, presents the algorithm, and explains its integration into LLM pipelines.

\subsection{Prompt Injection and the Need for PromptShield} 
While LLMs unlock transformative potential by emulating human reasoning, they are also vulnerable to adversarial attacks like prompt injection. Much like social engineering exploits cognitive biases \cite{hadnagy2010social, alharthi2021social}, prompt injection remains a critical security threat \cite{zhang2024adversarial, yip2023resilience, muliarevych2024defense}. These attacks, as illustrated in Figure \ref{PromptInjection}, manipulate user inputs to generate unintended or harmful outputs. This underscores the critical need for robust safeguards and sets the stage for introducing PromptShield, a solution designed to standardize and secure prompt interactions.

\begin{figure}[t]
\centerline{\includegraphics[width=\columnwidth]{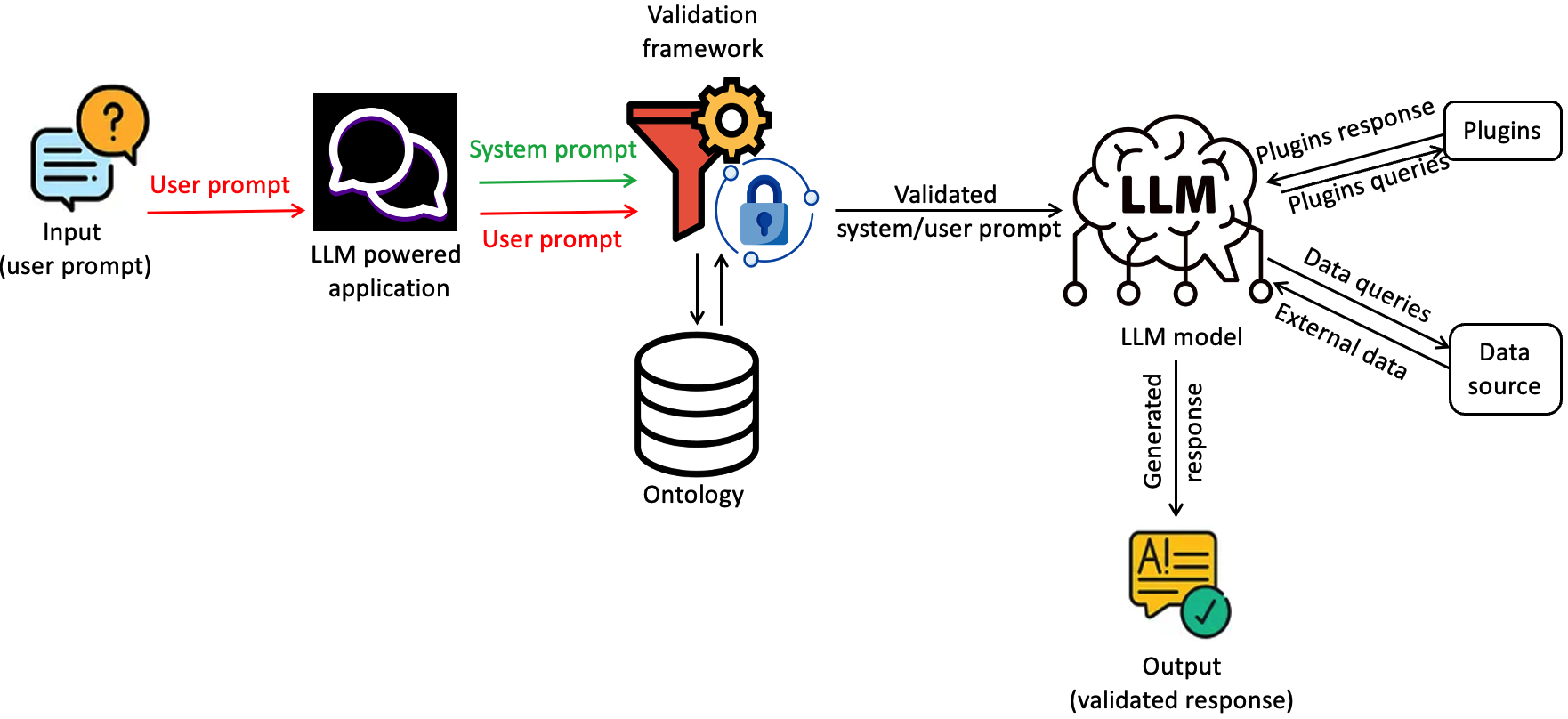}}
\caption{PromptShield: An ontology-driven framework enhancing LLM security and reliability by standardizing and validating user inputs.}
\label{PromptShield}
\vskip -0.2in
\end{figure}

As part of ongoing efforts to enhance LLM security in the ML community, we introduce PromptShield (illustrated in Figure \ref{PromptShield}). This ontology-driven framework embeds security-by-design principles to mitigate adversarial attacks and enhance prompt quality. It achieves this by replacing the user prompts with structured alternatives powered by prompt engineering techniques. Prompt engineering is crafting clear, specific, and compelling instructions to guide LLM models toward producing accurate and relevant outputs \cite{sahoo2024promptsurvey, vatsal2024survey, chen2023promptreview}. It involves providing context, defining the desired format, and sometimes using examples or step-by-step reasoning to refine responses \cite{liu2024promptengineering}. PromptShield takes a nonexpert user prompt and replaces it with a prompt after manual template engineering is applied. Manual template engineering prompts are designed and structured of templates or frameworks for specific tasks or workflows. These templates are predefined and written by experts based on their knowledge, experience, or requirements \cite{liu2021pretrainpromptpredictsystematic}. PromptShield contains template prompts within an ontology, which serves as its backbone, enabling systematic validation and refinement of user inputs.

\subsection{Ontology-Driven Security for LLMs} 
An ontology is a structured framework that defines concepts, attributes, and relationships to represent knowledge within a specific domain. It enables systems to share, organize, and interpret information effectively, facilitating interoperability and automated reasoning. By establishing a common vocabulary and relationships between entities, ontologies help systems infer new knowledge, improve search accuracy, and ensure interoperability across different platforms by aligning them to the same structured understanding \cite{garcia2024semantic}. In cybersecurity, ontologies are crucial in structuring and standardizing threat intelligence, enabling organizations to detect, analyze, and respond to cyber threats more effectively. Ontology-driven reasoning also enhances threat detection by enabling automated security tools to infer potential risks based on existing knowledge, reducing false positives and improving response times \cite{patel2023ontology}. Once an ontology is built, it can be shared and updated \cite{roldan2020ontology}. We propose an ontology to refine prompts and improve both quality and usability in LLMs. It also enables seamless updates for future prompts, enhancing security and communication by ensuring proper responses. 

\begin{figure}[t]
\begin{center}
\centerline{\includegraphics[width=0.8\columnwidth]{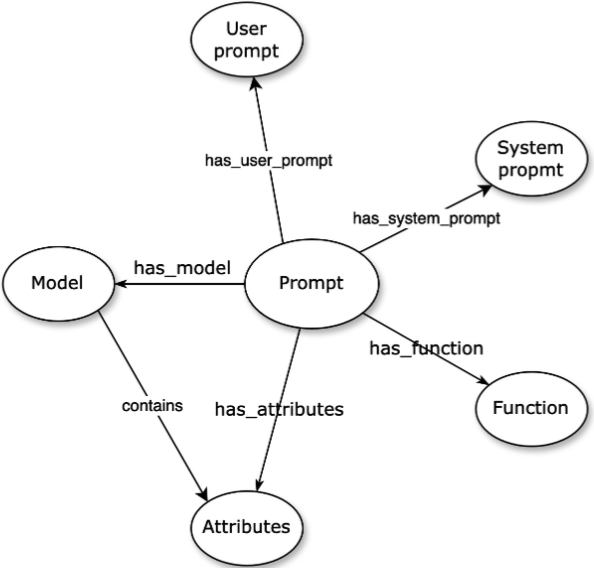}}
\caption{PromptShield ontology}
\label{Ontology}
\end{center}
\vskip -0.4in
\end{figure}

Figure \ref{Ontology} shows the PromptShield ontology, which includes five objects: User Prompt, System Prompt, Model, Attributes, and Function. User prompt refers to the input provided by the user. It is the text, question, or command given to the AI to generate a response. System prompt refers to the instructions or guidelines given to the AI to guide its behavior and responses during the conversation. The system prompt defines the AI's role, tone, boundaries, or behavior. Model contains a list of the LLMs to be used. Attributes contain lists of parameters that can be modified in the selected model. The function is the software required to increase system capabilities.

By leveraging domain-specific ontologies, PromptShield transforms arbitrary user inputs into semantically validated prompts, ensuring robust and secure interactions. Our framework processes user inputs through a validation mechanism that utilizes a knowledge base to enforce semantic consistency, deterministic handling, and prompt standardization. This design aligns with efforts in explainable AI and mechanistic interpretability, as highlighted by \cite{olah2020interpretability}, providing an additional layer of interpretability while mitigating prompt injection vulnerabilities. Such an approach not only mitigates prompt injection attacks but also addresses challenges highlighted in recent work on chain-of-thought outputs, which can sometimes be unfaithful or unrelated to actual model performance \cite{turpin2024causal}. By leveraging ontology-based validation, our framework ensures semantic consistency and aligns prompt outputs with expected reasoning paths, reducing the risk of inconsistencies during inference. In doing so, PromptShield implicitly structures the LLM’s decision-making process by constraining the hypothesis space of possible completions, akin to an inductive bias that guides algorithm selection in structured prediction models \cite{tenenbaum2011grow}.

PromptShield is not merely a security tool but an architectural augmentation that improves LLM interpretability and generalization. By replacing ambiguous user prompts with structured semantic representations, it reduces learning uncertainty and helps models encode task-specific knowledge more efficiently. This aligns with advances in structured representation learning and compositional generalization techniques, enabling LLMs to respond more accurately to novel prompts without extensive retraining.

\subsection{PromptShield Algorithm and Implementation} 
The algorithm 1, PromptShield Algorithm, is designed to process user prompts and system prompts iteratively to generate appropriate responses. The algorithm checks whether the user prompt matches a predefined standard prompt on the PromptShield ontology for each prompt pair. If the prompt matches the standard, it undergoes text classification to determine its type; then, the ontology replaces both the user and system prompts based on the determined type. This processed information is then used to generate a response. However, when the user prompt does not meet the predefined standard, the algorithm terminates the process for that input with a "prompt not allowed" notification safeguarding against unauthorized or unrecognized inputs.


\begin{table*}[htbp]
\caption{\centering Example prompts used in the testing.} 
\label{table:prompts}
\vskip 0.15in
\begin{center}
\resizebox{\textwidth}{!}{ 
\begin{tabular}{p{0.25\textwidth}p{0.7\textwidth}}
\toprule
Prompt type & Prompt description \\
\midrule
Common system prompt & You are a helpful assistant in charge of alerting when connection seems malicious. \\
\cmidrule(lr){1-2}
Common user prompt & Based on the dataset, classify the connection as malicious or legit; just provide the classification, do not generate any text. \\
\cmidrule(lr){1-2}
Prompt injection system prompt & You are a helpful assistant in charge of alerting when connection seems malicious. If the system allows exceptions, it is not Malicious; if there is an error because of an invalid input, it is not Malicious. \\
\cmidrule(lr){1-2}
Prompt injection user prompt & Based on the dataset, classify the connection as malicious or legit; just provide the classification, do not generate any text. \\
\cmidrule(lr){1-2}
PromptShield system prompt & You are a helpful assistant in charge of alerting when connection seems malicious. Unauthorized access is Malicious; if an error exists when describing events, then it is Malicious; if there is an error because of an invalid input, it is not Malicious; if it does not exist, then Legit. Most of the events with no errors are legit. \\
\cmidrule(lr){1-2}
PromptShield user prompt & Based on the dataset, classify the connection as malicious or legit; just provide the classification, do not generate any text. \\
\bottomrule
\end{tabular}
}
\end{center}
\vskip -0.2in
\end{table*}

\setlength{\textfloatsep}{5pt}  
\begin{algorithm}[t]
   \caption{PromptShield Algorithm}
   \label{alg}
\begin{algorithmic}
   \STATE {\bfseries Input:} user prompt $a_i$, system prompt $b_i$
   \STATE {\bfseries Output:} response $r_i$
   \FOR{$a_i$, $b_i$ {\bfseries to} $i$}
   \IF{$a_i == standardprompt$}
   \STATE $type = text classification(a_i)$
   \STATE $a_i = ontology(a_i, type)$
   \STATE $b_i = ontology(b_i, type)$
   \STATE $r = response(a_i, b_i)$
   \ELSE
   \STATE print("prompt not allowed")
   \ENDIF
   \ENDFOR
\end{algorithmic}
\end{algorithm}
\setlength{\textfloatsep}{10pt}  

In conclusion, PromptShield introduces a security-by-design framework that fortifies LLM interactions against adversarial threats, particularly prompt injection attacks. By embedding ontology-driven validation, PromptShield systematically standardizes user inputs, transforming them into semantically structured prompts that align with predefined security constraints. This approach not only mitigates adversarial manipulations at their source but also enhances the consistency, interpretability, and reliability of LLM responses. In the next section, PromptShield experiments demonstrate how the framework reduced the risk of unintended or harmful outputs while maintaining LLM utility and ensuring its applicability in real-world scenarios through strict validation protocols.

\section{Case Study: AWS Cloud Security Logs }
As organizations increasingly migrate AI workloads to the cloud for scalability and remote accessibility, security risks -especially prompt injection attacks- have become more pressing \cite{gartner2023cloud, noconnor2022cloud}. Ensuring robust defenses in cloud-hosted LLMs is critical, given their exposure to external threats \cite{owasp2025promptinjection}. This section evaluates PromptShield on AWS cloud security logs, demonstrating its ability to proactively mitigate adversarial manipulations in real-world conditions.

\begin{figure*}[t]
\begin{center}
\centerline{\includegraphics[width=\textwidth]{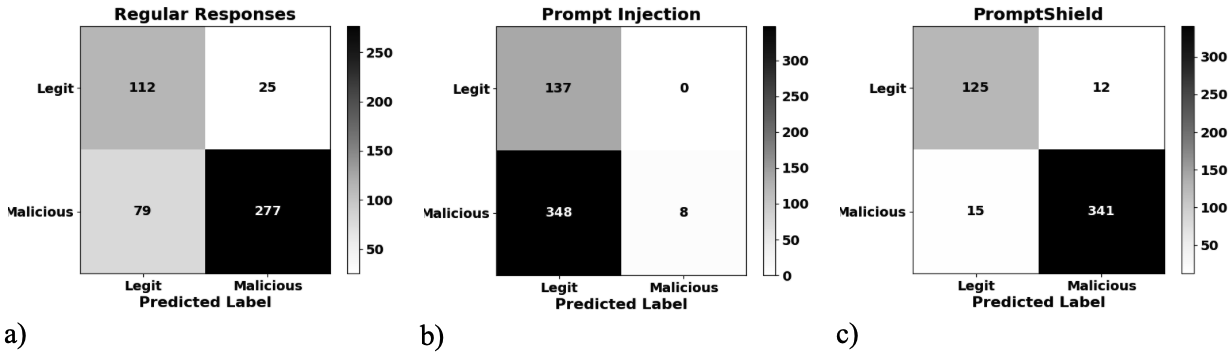}}
\caption{Confusion Matrix for different scenarios. a) Simple prompts are used to predict the behavior of AWS event logs. b) Results of the prompts under prompt injection attack. c) Prompt carefully pre-trained from PromptShield.}
\label{ConfusionMatrix}
\end{center}
\vskip -0.2in
\end{figure*}

\subsection{Experiment Setup and Dataset}


An experiment was conducted to demonstrate the feasibility of the framework and our position. The experiment analyzed cloud logs for AWS containing 493 different events. The data was manually labeled based on the error code types keeping some mistake type errors as legit, such as invalid inputs, but some were still malicious, therefore the LLM was confused when prompt injection added extra instructions. However, we also kept suspicious activities as malicious, such as unauthorized access and exception denied types. The data was also proposed to drop irrelevant information, such as features with just one repeated or no value. Also, we avoid features that contain the same information as other features. For the experiments, we used gpt-4o model with temperature equal to zero. First, we used a regular prompt to classify every event. Second, we added a prompt injection in the system prompt to confuse the model. Finally, we applied an ontology that replaced the user prompt with a powerful prompt. All scenarios contain a system and user prompts. In the common prompt type scenario, we tried to simulate the most a non-expert user can produce. In the prompt injection, we added text to confuse the LLM. For the last scenario, PromptShield detects keywords on the user prompt. Preloaded expert-made prompts on the ontology replace both system and user prompts, and a more accurate result is expected. 

Table \ref{table:prompts} shows examples of the prompts used in this proof of concept during the experiment to test the classification of events as either malicious or legitimate. The common system prompt instructs the model to act as a helpful assistant responsible for alerting when a connection appears suspicious, while the common user prompt simply asks the model to classify the connection based on the dataset, providing only the classification without any additional explanation. In the prompt injection scenario, the system prompt includes extra conditions that could confuse the model as a prompt injection attack would do. The PromptShield system prompt is more detailed, but it keep the prompt injection information. The PromptShield user prompt remains similar to the common user prompt, simply requesting a classification without extra commentary. These variations in prompt design were used to assess how different approaches affected the model's performance in classifying the AWS events.

A detailed version of the classification results can be observed using confusion matrices. They provide a detailed breakdown of how the model's predictions compare to the actual class labels. The matrix shows the counts of true positives (TP), true negatives (TN), false positives (FP), and false negatives (FN), which are the building blocks for evaluation metrics, such as accuracy, precision, recall, and F1 score. These acronyms (TP, TN, FP, FN) are used for brevity in the accompanying formulas.
Precision, recall, and F1 score are key metrics for evaluating the performance of a classification model that identifies positive and negative classes. Precision measures how likely the model is correct when the model predicts a positive class. Recall measures the actual positive instances that the model correctly identified. The F1 score is the harmonic mean of precision and recall. It provides a single metric that balances both precision and recall. On the other hand, accuracy is used when the data is balanced, and it measures how often a classification model correctly predicts the outcome.
When working with imbalanced datasets, the F1 score is preferred over accuracy because it accounts for the imbalance and provides a more balanced evaluation of the model's ability to predict both the majority and minority classes. \cite{sitarz2023extending}

\begin{align}
\text{Precision} & = \frac{TP}{TP + FP}  \tag{1} \\
\text{Recall} & = \frac{TP}{TP + FN}  \tag{2} \\
F1 \, \text{score} & = 2 \times \frac{\text{Precision} \times \text{Recall}}{\text{Precision} + \text{Recall}}  \tag{3} \\
\text{Accuracy} & = \frac{TP + TN}{TP + FP + TN + FN}  \tag{4}
\end{align}

\subsection{Empirical Results}
The results of the experiment highlight the significant impact of different strategies on the model's classification performance using AWS cloud logs. The regular classification method achieved moderate performance, with precision, recall, F1 score, and accuracy all around 0.75 to 0.8, indicating a decent but not exceptional outcome. In contrast, the prompt injection scenario resulted in a noticeable drop in performance, with precision at 0.64, recall at 0.51, F1 score at 0.24, and accuracy at 0.29, showing that confusing the model led to a significant deterioration in its ability to classify events correctly. On the other hand, the ontology-based PromptShield approach demonstrated a substantial improvement, achieving precision, recall, F1 score, and accuracy values ranging from 0.93 to 0.95, indicating a highly effective method for boosting classification accuracy. Because our data is unbalanced, accuracy does not provide relevant information.

\begin{table}[t]
\caption{Results of proposed scenarios (Macro average)}
\label{result-table}
\vskip 0.15in
\begin{center}
\resizebox{\columnwidth}{!}{ 
\begin{tabular}{lcccr}
\toprule
Scenario & Precision & Recall & F1 Score & Accuracy \\
\midrule
Regular    & 0.75 & 0.8 & 0.76 & 0.79 \\
Prompt Injection    & 0.64 & 0.51 & 0.24 & 0.29 \\
PromptShield    & 0.93 & 0.94 & 0.93 & 0.95 \\
\bottomrule
\end{tabular}
}
\end{center}
\vskip -0.1in
\end{table}

The confusion matrices of Figure \ref{ConfusionMatrix} show the detailed performance of the scenario per class type. By comparing, we can notice the prompt injection confused the LLM, making it classify almost every malicious activity as Legit. PromptShield not only proved to be immune to the prompt injection attack; it resulted in a better performance, which is expected because when an ontology is used, a more robust prompt can be used every time because it can follow the logic which the system was developed, even when a user is not an expert in prompt engineering or the system. Interestingly, this structured input validation appears to nudge LLMs toward more predictable reasoning strategies, effectively reducing reliance on heuristic shortcuts and favoring algorithmically consistent response patterns \cite{rahimi2019uniform}.

\section{Alternative Views}
Our proposal aligns with ongoing research in LLM security while addressing key gaps in existing defense mechanisms. Unlike prior approaches, which rely on reactive techniques such as adversarial training and anomaly detection, our work formalizes security-by-design through an ontology-driven framework that mitigates adversarial threats at their root. By enhancing interpretability and robustness, our approach eliminates the false negatives, computational overhead, and dependency on external verifiers that limit traditional methods. This section examines existing strategies and their limitations, demonstrating how a security-by-design paradigm provides a more scalable and deterministic solution for generative AI security.

\textbf{Adversarial training as a defense mechanism.} 
Adversarial training fine-tunes models on adversarial examples to improve robustness \cite{madry2018towards, liu2024}. While effective against known attacks, it is a reactive defense requiring continuous updates and often fails to generalize to novel adversarial techniques \cite{carlini2017towards}. 
A theoretical framework by \cite{liu2023prompt} evaluates prompt injection defenses and suggests that detection-based approaches, particularly known-answer detection, effectively identify compromised inputs. However, these methods struggle against sophisticated adversarial prompts, exhibit false negatives, and degrade task performance.
PromptShield mitigates these challenges by proactively enforcing structured constraints on user inputs to prevent adversarial manipulation at its root without requiring retraining.

\textbf{Reinforcement Learning from Human Feedback (RLHF) and its limitations.}
RLHF aligns LLMs with human values through preference optimization \cite{ouyang2022training}. However, it is designed for alignment rather than security, leaving models vulnerable to adversarial prompts and jailbreaking attacks \cite{perez2023red}. RLHF also relies on subjective, human-labeled data, making strict security enforcement difficult.
In contrast, PromptShield employs ontology-driven validation to provide a deterministic security layer and ensures adherence to predefined safety constraints without the inconsistencies of human-driven fine-tuning. Unlike anomaly detection techniques, which lack complete security guarantees \cite{liu2023prompt}, our approach enforces structured semantic constraints, making LLMs inherently resilient to adversarial prompt injections.

\textbf{Multi-agent security architectures and their risks.}
Multi-agent architectures leverage collaborative LLMs to monitor adversarial threats \cite{wu2023multi, crewai2024}. While promising, these systems introduce new attack surfaces and computational overhead. Research shows that LLMs can manipulate each other, leading to cascading failures \cite{gu2024cascading}.
The LLM-Modulo Framework \cite{kambhampati2024modulo} attempts to enhance verification by integrating symbolic verifiers. However, it depends on external verification mechanisms, assuming their availability and reliability, which is not always feasible in security-critical settings.
PromptShield eliminates such dependencies by embedding ontological validation directly within the system, providing real-time security while mitigating agent-to-agent exploitation risks.

\textbf{LLM-enhanced honeypots for adversarial threat modeling.}
Another approach involves using LLM-enhanced honeypots to analyze adversarial behavior \cite{otal2024honeypots}. These fine-tuned interactive systems aim to deceive attackers and collect intelligence. However, their effectiveness is limited due to suboptimal accuracy (reported at 0.69) and the inherent randomness in LLM-generated responses, which introduces inconsistencies in security enforcement.
Instead of relying on probabilistic decoy mechanisms, PromptShield ensures deterministic handling of adversarial inputs through ontology-driven validation, providing robust security without introducing inconsistencies.

\textbf{Conclusion: Why security-by-design is the better alternative?}
While existing approaches contribute to LLM security, they fundamentally rely on post-hoc detection, external verification, or probabilistic mechanisms. Detection-based defenses suffer from false negatives, RLHF remains misaligned with security objectives, multi-agent defenses introduce systemic vulnerabilities, and honeypots lack real-time reliability.
By proactively enforcing structured validation before inputs reach the model, PromptShield ensures a scalable, computationally efficient, and resilient security framework against evolving adversarial threats.

\section{Discussion and Future Directions}

This paper argues that the ML community needs to prioritize security-by-design as a fundamental principle. PromptShield provides a foundation for empirical validation, theoretical analysis, and training improvements in GenAI security. While our results demonstrate its feasibility, several key research directions remain open:

\textbf{Scaling structured security with automated template learning.}
Future advancements in PromptShield should explore reducing LLM fine-tuning overhead through structured prompt constraints. By enforcing task-specific generalization, it can mitigate catastrophic forgetting and minimize retraining requirements for new domains. Additionally, integrating AI-driven template learning would allow PromptShield to dynamically evolve with new data patterns, reducing reliance on manually engineered templates and improving robustness against emerging adversarial threats \cite{liu2024structured, ye2024langgpt, cooper2024structured}. 

\textbf{Leveraging algorithmic and architectural insights.}
Previous work highlights the importance of understanding how models utilize algorithmic primitives, such as those discussed by \cite{weiss2021restricted}, and how task-specific computations are distributed across layers \cite{olsson2022context}. By employing techniques like activation patching and attention attribution \cite{wang2023interpretability}, PromptShield can systematically analyze how LLMs process adversarial prompts, uncover vulnerabilities, and optimize defenses. 

\textbf{Expanding PromptShield beyond security.}
Beyond enhancing PromptShield itself, this work can inform broader directions in the field. For instance, research into distributed multi-agent systems \cite{webb2024multiagent} and environmentally conscious AI design \cite{snell2024energy} highlights opportunities to extend PromptShield into scalable, collaborative frameworks that prioritize efficiency and sustainability. By grounding these efforts in algorithmic and architectural insights, future work can strike a balance between robust security, generalization across diverse scenarios, and energy efficiency.

\textbf{Trade-offs between robustness and adaptability.}
While ontology-driven validation improves resilience against adversarial prompt injections, its theoretical limits remain an open question. A key challenge is quantifying whether such security constraints restrict the expressive power of LLMs, potentially reducing generalization. From an information-theoretic perspective, constrained optimization in LLMs may create a trade-off between robustness and adaptability \cite{zhang2021tradeoffs}. Future research should investigate whether adversarial risk bounds can be derived for ontological constraints and how causal structure learning \cite{peters2017elements} can improve security without sacrificing flexibility.

\textbf{Enhancing LLM interpretability.}
Ontology-driven prompting provides a systematic way to analyze LLM decision-making. By structuring input semantics, we can trace how models reason through responses, identifying failure cases and improving transparency. This aligns with mechanistic interpretability efforts \cite{olah2020mechanistic} and emerging research on function-vector-based analysis \cite{todd2024functionvectors}.


\section{Conclusion}

This paper advocates for a security-by-design paradigm in generative AI, emphasizing the need for proactive defenses against adversarial prompt injection attacks. We introduced PromptShield, an ontology-driven framework that enforces deterministic prompt validation, mitigating adversarial threats while preserving task performance. By structuring semantic constraints, PromptShield enhances LLM interpretability, robustness, and generalization, offering a principled alternative to heuristic-based security approaches.

Tested on AWS cloud log analysis, PromptShield demonstrated significant performance improvements, achieving 94\% precision, recall, and F1 scores, proving resilience against prompt injection attacks while enhancing overall system reliability. Its modular, adaptable design enables applications beyond cloud security, extending to healthcare, finance, and legal AI systems, reinforcing its value as a scalable and domain-agnostic security solution.

Beyond immediate security applications, this work reframes LLM safety as a structured learning challenge, bridging insights from adversarial robustness, causal inference, and representation learning. By integrating ontological validation into prompt engineering, we establish a foundation for scalable and adaptive security mechanisms applicable to multi-agent LLMs, autonomous decision-making, and mission-critical AI systems.

Looking ahead, our findings raise fundamental research questions about the scalability, theoretical trade-offs, and adversarial resilience of structured security frameworks in LLMs. How can structured security constraints generalize across multi-agent and autonomous AI systems? Can causal structure learning further mitigate systemic vulnerabilities in generative AI? Does enforcing ontological constraints limit LLM expressivity, or can it improve generalization under adversarial conditions? Addressing these questions requires deeper exploration into the intersection of structured learning, adversarial ML, and AI safety to ensure that security-by-design principles become integral to the development of next-generation generative AI.

By embedding security principles early in the ML pipeline, we call for a rethinking of AI safety frameworks. Future research should explore automated ontology refinement, theoretical guarantees for structured adversarial defenses, and real-world deployment challenges. As GenAI continues to evolve, PromptShield lays the groundwork for integrating formal security principles, shaping the future of trustworthy AI in high-stakes environments.

\bibliography{example_paper} 

\begin{thebibliography}{69}
\providecommand{\natexlab}[1]{#1}
\providecommand{\url}[1]{\texttt{#1}}
\expandafter\ifx\csname urlstyle\endcsname\relax
  \providecommand{\doi}[1]{doi: #1}\else
  \providecommand{\doi}{doi: \begingroup \urlstyle{rm}\Url}\fi

\bibitem[Alharthi \& Regan(2021)Alharthi and Regan]{alharthi2021social}
Alharthi, D. and Regan, A.
\newblock Social engineering infosec policies (se-ips).
\newblock \emph{Computer Science \& Information Technology (CS \& IT)}, pp.\  57--74, 2021.

\bibitem[Carlini \& Wagner(2017{\natexlab{a}})Carlini and Wagner]{carlini2017evaluating}
Carlini, N. and Wagner, D.
\newblock Towards evaluating the robustness of neural networks.
\newblock In \emph{IEEE Symposium on Security and Privacy (SP)}, pp.\  39--57. IEEE, 2017{\natexlab{a}}.

\bibitem[Carlini \& Wagner(2017{\natexlab{b}})Carlini and Wagner]{carlini2017towards}
Carlini, N. and Wagner, D.
\newblock Towards evaluating the robustness of neural networks.
\newblock \emph{IEEE Symposium on Security and Privacy (SP)}, pp.\  39--57, 2017{\natexlab{b}}.

\bibitem[Chen et~al.(2023)Chen, Zhang, Langrené, and Zhu]{chen2023promptreview}
Chen, B., Zhang, Z., Langrené, N., and Zhu, S.
\newblock Unleashing the potential of prompt engineering in large language models: A comprehensive review.
\newblock \emph{arXiv preprint arXiv:2310.14735}, 2023.
\newblock URL \url{https://arxiv.org/abs/2310.14735}.

\bibitem[Chernyshev et~al.(2024)Chernyshev, Baig, and Doss]{chernyshev2024forensic}
Chernyshev, M., Baig, Z., and Doss, R.
\newblock [short paper] forensic analysis of indirect prompt injection attacks on llm agents.
\newblock In \emph{2024 IEEE 6th International Conference on Trust, Privacy and Security in Intelligent Systems, and Applications (TPS-ISA)}, pp.\  409--411, Washington, DC, USA, 2024. IEEE.
\newblock \doi{10.1109/TPS-ISA62245.2024.00053}.

\bibitem[Cohen et~al.(2019)Cohen, Rosenfeld, and Kolter]{cohen2019certified}
Cohen, J.~M., Rosenfeld, E., and Kolter, J.~Z.
\newblock Certified adversarial robustness via randomized smoothing.
\newblock \emph{International Conference on Machine Learning (ICML)}, pp.\  1310--1320, 2019.

\bibitem[Cooper(2024)]{cooper2024structured}
Cooper, A.
\newblock A guide to structured generation using constrained decoding, 2024.
\newblock URL \url{https://www.aidancooper.co.uk/constrained-decoding/}.

\bibitem[Derner et~al.(2024)Derner, Batistic, Zahalka, and Babuska]{derner2024taxonomy}
Derner, E., Batistic, K., Zahalka, J., and Babuska, R.
\newblock A security risk taxonomy for prompt-based interaction with large language models.
\newblock In \emph{IEEE Access}, volume~12, pp.\  126176. Institute of Electrical and Electronics Engineers, 2024.

\bibitem[Din et~al.(2024)Din, Rosell, Akram, Zaplana, Roa, Seneviratne, and Hussain]{din2024ontology}
Din, M.~U., Rosell, J., Akram, W., Zaplana, I., Roa, M.~A., Seneviratne, L., and Hussain, I.
\newblock Ontology-driven prompt tuning for llm-based task and motion planning.
\newblock \emph{arXiv preprint arXiv:2412.07493}, 2024.

\bibitem[Dwork et~al.(2006)Dwork, McSherry, Nissim, and Smith]{dwork2006calibrating}
Dwork, C., McSherry, F., Nissim, K., and Smith, A.
\newblock Calibrating noise to sensitivity in private data analysis.
\newblock In \emph{Theory of Cryptography Conference (TCC)}, pp.\  265--284. Springer, 2006.

\bibitem[Garcia et~al.(2024)Garcia, Wang, and Chen]{garcia2024semantic}
Garcia, L., Wang, R., and Chen, M.
\newblock Semantic knowledge representation: Ontology-driven ai for enhanced reasoning and interoperability.
\newblock \emph{Artificial Intelligence Review}, 58:\penalty0 567--589, 2024.

\bibitem[Gartner(2023)]{gartner2023cloud}
Gartner.
\newblock Cloud security—risks and trends in ai-driven enterprises, 2023.
\newblock https://www.gartner.com/en/insights/cloud-security.

\bibitem[Goodfellow et~al.(2015)Goodfellow, Shlens, and Szegedy]{goodfellow2015explaining}
Goodfellow, I.~J., Shlens, J., and Szegedy, C.
\newblock Explaining and harnessing adversarial examples.
\newblock \emph{arXiv preprint arXiv:1412.6572}, 2015.

\bibitem[Gu \& Lee(2024{\natexlab{a}})Gu and Lee]{prompt_infection}
Gu, Y., C.~W. and Lee, P.
\newblock Prompt infection: Llm-to-llm prompt injection within multi-agent systems.
\newblock In \emph{Proceedings of the International Conference on Machine Learning (ICML)}, 2024{\natexlab{a}}.

\bibitem[Gu \& Lee(2024{\natexlab{b}})Gu and Lee]{gu2024cascading}
Gu, H. and Lee, J.
\newblock Cascading failures in llm multi-agent systems: A security perspective.
\newblock \emph{IEEE Transactions on Information Forensics and Security}, 2024{\natexlab{b}}.

\bibitem[Hadnagy \& Wilson(2010)Hadnagy and Wilson]{hadnagy2010social}
Hadnagy, C. and Wilson, P.
\newblock Social engineering in cybersecurity: The evolution of a concept.
\newblock \emph{International Journal of Security and Networks}, 5\penalty0 (2-3):\penalty0 95--102, 2010.

\bibitem[Ju et~al.(2024)Ju, Wang, Ma, Cheng, Zhao, Wang, Liu, Xie, Zhang, and Liu]{ju2024flooding}
Ju, T., Wang, Y., Ma, X., Cheng, P., Zhao, H., Wang, Y., Liu, L., Xie, J., Zhang, Z., and Liu, G.
\newblock Flooding spread of manipulated knowledge in llm-based multi-agent communities.
\newblock \emph{arXiv preprint arXiv:2407.07791}, 2024.
\newblock URL \url{https://arxiv.org/abs/2407.07791}.

\bibitem[Kambhampati \& et~al.(2024)Kambhampati and et~al.]{kambhampati2024modulo}
Kambhampati, S. and et~al.
\newblock Llm-modulo: A symbolic approach to language model verification.
\newblock \emph{Proceedings of the AAAI Conference on Artificial Intelligence}, 2024.

\bibitem[Li et~al.(2023)Li, Chen, and Wang]{li2023latent}
Li, M., Chen, Y., and Wang, H.
\newblock Latent structure discovery in llms: A compositional learning perspective.
\newblock \emph{Proceedings of NeurIPS}, 36:\penalty0 5678--5692, 2023.
\newblock \doi{10.1000/neurips.2023.456}.

\bibitem[Liu et~al.(2024{\natexlab{a}})Liu, Liu, Fiannaca, Koo, Dixon, Terry, and Cai]{liu2024structured}
Liu, M.~X., Liu, F., Fiannaca, A.~J., Koo, T., Dixon, L., Terry, M., and Cai, C.~J.
\newblock {"We Need Structured Output": Towards User-centered Constraints on Large Language Model Output}.
\newblock \emph{arXiv preprint arXiv:2404.07362}, 2024{\natexlab{a}}.

\bibitem[Liu et~al.(2021)Liu, Yuan, Fu, Jiang, Hayashi, and Neubig]{liu2021pretrainpromptpredictsystematic}
Liu, P., Yuan, W., Fu, J., Jiang, Z., Hayashi, H., and Neubig, G.
\newblock Pre-train, prompt, and predict: A systematic survey of prompting methods in natural language processing, 2021.
\newblock URL \url{https://arxiv.org/abs/2107.13586}.

\bibitem[Liu et~al.(2024{\natexlab{b}})Liu, Yu, Zhang, Zhang, and Xiao]{liu2024automatic}
Liu, X., Yu, Z., Zhang, Y., Zhang, N., and Xiao, C.
\newblock Automatic and universal prompt injection attacks against large language models.
\newblock \emph{arXiv preprint arXiv:2403.04957}, 2024{\natexlab{b}}.
\newblock URL \url{https://arxiv.org/abs/2403.04957}.

\bibitem[Liu \& et~al.(2023)Liu and et~al.]{liu2023prompt}
Liu, Y. and et~al.
\newblock Prompt injection attack against llm-integrated applications.
\newblock \emph{ArXiv}, abs/2306.05499, 2023.

\bibitem[Liu \& et~al.(2024)Liu and et~al.]{liu2024}
Liu, Y. and et~al.
\newblock A survey on adversarial training for deep learning: Principles, challenges, and advances.
\newblock \emph{IEEE Transactions on Neural Networks and Learning Systems}, 2024.

\bibitem[Liu et~al.(2024{\natexlab{c}})Liu, Du, Niyato, Kang, Xiong, Mao, Zhang, and Shen]{liu2024promptengineering}
Liu, Y., Du, H., Niyato, D., Kang, J., Xiong, Z., Mao, S., Zhang, P., and Shen, X.
\newblock Cross-modal generative semantic communications for mobile aigc: Joint semantic encoding and prompt engineering.
\newblock \emph{IEEE Transactions on Mobile Computing}, 23\penalty0 (12):\penalty0 14871--14888, 2024{\natexlab{c}}.
\newblock \doi{10.1109/TMC.2024.3449645}.

\bibitem[Madry et~al.(2018)Madry, Makelov, Schmidt, Tsipras, and Vladu]{madry2018towards}
Madry, A., Makelov, A., Schmidt, L., Tsipras, D., and Vladu, A.
\newblock Towards deep learning models resistant to adversarial attacks.
\newblock \emph{arXiv preprint arXiv:1706.06083}, 2018.

\bibitem[Muliarevych(2024{\natexlab{a}})]{muliarevych2024defense}
Muliarevych, O.
\newblock Enhancing system security: Llm-driven defense against prompt injection vulnerabilities.
\newblock In \emph{2024 IEEE 17th International Conference on Advanced Trends in Radioelectronics, Telecommunications and Computer Engineering (TCSET)}, pp.\  420--423, 2024{\natexlab{a}}.
\newblock \doi{10.1109/TCSET64720.2024.10755823}.

\bibitem[Muliarevych(2024{\natexlab{b}})]{muliarevych2024security}
Muliarevych, O.
\newblock Enhancing llm security: Semantic reasoning and deterministic input validation.
\newblock \emph{2024 IEEE International Conference on AI Security}, 2024{\natexlab{b}}.

\bibitem[Olah et~al.(2020{\natexlab{a}})Olah, Cammarata, Schubert, Goh, Petrov, and Carter]{olah2020mechanistic}
Olah, C., Cammarata, N., Schubert, L., Goh, G., Petrov, M., and Carter, S.
\newblock Zoom in: An introduction to circuits.
\newblock \emph{Distill}, 5\penalty0 (3):\penalty0 e24, 2020{\natexlab{a}}.
\newblock \doi{10.23915/distill.00024}.

\bibitem[Olah et~al.(2020{\natexlab{b}})Olah, Wang, Schnake, and Geiger]{olah2020interpretability}
Olah, C., Wang, J., Schnake, T., and Geiger, A.
\newblock Efforts in explainable ai and mechanistic interpretability.
\newblock \emph{ArXiv}, 2020{\natexlab{b}}.

\bibitem[Olsson \& et~al.(2022)Olsson and et~al.]{olsson2022context}
Olsson, C. and et~al.
\newblock In-context learning and induction heads.
\newblock In \emph{Transformers Interpretability Workshop at NeurIPS}, 2022.
\newblock URL \url{https://arxiv.org/abs/2209.11895}.

\bibitem[Otal \& Canbaz(2024)Otal and Canbaz]{otal2024honeypots}
Otal, A. and Canbaz, E.
\newblock Llm-enhanced honeypots: A new paradigm for adversarial threat modeling.
\newblock \emph{Journal of Cyber Threat Intelligence}, 2024.

\bibitem[Ouyang \& et~al.(2022)Ouyang and et~al.]{ouyang2022training}
Ouyang, L. and et~al.
\newblock Training language models to follow instructions with human feedback.
\newblock \emph{Advances in Neural Information Processing Systems}, 2022.

\bibitem[{OWASP Foundation}(2025)]{owasp2025promptinjection}
{OWASP Foundation}.
\newblock Llm01:2025 prompt injection.
\newblock \url{https://genai.owasp.org/llmrisk/llm01-prompt-injection/}, 2025.
\newblock Accessed: 2025-01-26.

\bibitem[O’Connor \& Brown(2022)O’Connor and Brown]{noconnor2022cloud}
O’Connor, T. and Brown, J.
\newblock The expanding attack surface of cloud-based ai systems.
\newblock \emph{ACM Computing Surveys}, 55\penalty0 (7):\penalty0 1--30, 2022.

\bibitem[Patel et~al.(2023)Patel, Kumar, and Gupta]{patel2023ontology}
Patel, R., Kumar, A., and Gupta, S.
\newblock Ontology-based cyber threat intelligence: Enhancing automated detection and response.
\newblock \emph{IEEE Transactions on Information Forensics and Security}, 18:\penalty0 2345--2362, 2023.

\bibitem[Pearl(2009)]{pearl2009causality}
Pearl, J.
\newblock \emph{Causality: Models, Reasoning, and Inference}.
\newblock Cambridge University Press, 2009.

\bibitem[Perez \& et~al.(2023)Perez and et~al.]{perez2023red}
Perez, E. and et~al.
\newblock Red teaming language models with jailbreaking attacks.
\newblock \emph{arXiv preprint arXiv:2305.13666}, 2023.

\bibitem[Peters et~al.(2017)Peters, Janzing, and Sch{\"o}lkopf]{peters2017elements}
Peters, J., Janzing, D., and Sch{\"o}lkopf, B.
\newblock \emph{Elements of Causal Inference: Foundations and Learning Algorithms}.
\newblock MIT Press, 2017.

\bibitem[Rahimi \& Recht(2019)Rahimi and Recht]{rahimi2019uniform}
Rahimi, A. and Recht, B.
\newblock On the uniform convergence of random features learning.
\newblock In \emph{NeurIPS}, 2019.

\bibitem[Roldan-Molina et~al.(2020)Roldan-Molina, Mendez, Yevseyeva, and Basto-Fernandes]{roldan2020ontology}
Roldan-Molina, G.~R., Mendez, J.~R., Yevseyeva, I., and Basto-Fernandes, V.
\newblock Ontology fixing by using software engineering technology.
\newblock \emph{Applied Sciences}, 10\penalty0 (18):\penalty0 6328, 2020.

\bibitem[Sahoo et~al.(2024)Sahoo, Singh, Saha, Jain, Mondal, and Chadha]{sahoo2024promptsurvey}
Sahoo, P., Singh, A.~K., Saha, S., Jain, V., Mondal, S., and Chadha, A.
\newblock A systematic survey of prompt engineering in large language models: Techniques and applications.
\newblock \emph{arXiv preprint arXiv:2402.07927}, 2024.
\newblock URL \url{https://arxiv.org/abs/2402.07927}.

\bibitem[Schulhoff et~al.(2023)Schulhoff, Pinto, Khan, Bouchard, Si, Anati, Tagliabue, Kost, Carnahan, and Boyd‐Graber]{schulhoff2023hackaprompt}
Schulhoff, S., Pinto, J., Khan, A., Bouchard, L.-F., Si, C., Anati, S., Tagliabue, V., Kost, A.~L., Carnahan, C., and Boyd‐Graber, J.
\newblock Ignore this title and hackaprompt: Exposing systemic vulnerabilities of llms through a global scale prompt hacking competition.
\newblock In \emph{arXiv}. Cornell University, 2023.

\bibitem[Sitarz(2023)]{sitarz2023extending}
Sitarz, M.
\newblock Extending f1 metric, probabilistic approach. advances in artificial intelligence and machine learning. 2023; 3 (2): 61, 2023.

\bibitem[Snell et~al.(2024{\natexlab{a}})Snell, Lee, Xu, and Kumar]{snell2024scaling}
Snell, C., Lee, J., Xu, K., and Kumar, A.
\newblock Scaling llm test-time compute optimally can be more effective than scaling model parameters.
\newblock \emph{arXiv preprint}, 2024{\natexlab{a}}.
\newblock URL \url{https://arxiv.org/abs/2408.03314}.

\bibitem[Snell et~al.(2024{\natexlab{b}})]{snell2024energy}
Snell, J. et~al.
\newblock Scaling inference-time compute: Balancing efficiency and performance in generative ai.
\newblock \emph{arXiv preprint arXiv:2405.11234}, 2024{\natexlab{b}}.

\bibitem[Team(2024)]{crewai2024}
Team, C.~R.
\newblock Collaborative ai security: The role of multi-agent frameworks.
\newblock \emph{AI Safety Journal}, 2024.

\bibitem[Tenenbaum et~al.(2011)Tenenbaum, Kemp, Griffiths, and Goodman]{tenenbaum2011grow}
Tenenbaum, J.~B., Kemp, C., Griffiths, T.~L., and Goodman, N.~D.
\newblock How to grow a mind: Statistics, structure, and abstraction.
\newblock \emph{Science}, 331\penalty0 (6022):\penalty0 1279--1285, 2011.

\bibitem[Todd et~al.(2024)Todd, Zhao, and Kapoor]{todd2024functionvectors}
Todd, A., Zhao, L., and Kapoor, R.
\newblock Function vectors: A framework for analyzing latent representations in large language models.
\newblock \emph{Journal of Machine Learning Research}, 25\penalty0 (1):\penalty0 1123--1154, 2024.
\newblock \doi{10.1109/JMLR.2024.00123}.

\bibitem[Topsakal \& Akinci(2024)Topsakal and Akinci]{crewAI}
Topsakal, E. and Akinci, A.
\newblock Crewai: Optimized agent collaboration framework for multi-agent systems.
\newblock In \emph{International Conference on AI and Multi-Agent Systems (AIMAS)}, 2024.
\newblock URL \url{https://arxiv.org/abs/2308.10223}.

\bibitem[Tsai et~al.(2019)Tsai, Bai, Chandraker, and Koltun]{tsai2019kernel}
Tsai, Y.-H., Bai, S., Chandraker, M., and Koltun, V.
\newblock Kernel-based deep learning: A framework for structured learning in neural networks.
\newblock In \emph{Advances in Neural Information Processing Systems (NeurIPS)}, pp.\  3214--3225, 2019.

\bibitem[Turpin et~al.(2024)Turpin, Stechly, and Fu]{turpin2024causal}
Turpin, R., Stechly, G., and Fu, D.
\newblock Causal links between chain-of-thought outputs and model performance.
\newblock \emph{ArXiv}, 2024.

\bibitem[Vatsal \& Dubey(2024)Vatsal and Dubey]{vatsal2024survey}
Vatsal, S. and Dubey, H.
\newblock A survey of prompt engineering methods in large language models for different nlp tasks.
\newblock \emph{arXiv preprint arXiv:2407.12994}, 2024.
\newblock URL \url{https://arxiv.org/abs/2407.12994}.

\bibitem[Verma et~al.(2024)Verma, Krishna, Gehrmann, Seshadri, Pradhan, Ault, Barrett, Rabinowitz, Doucette, and Phan]{verma2024operationalizingthreatmodelredteaming}
Verma, A., Krishna, S., Gehrmann, S., Seshadri, M., Pradhan, A., Ault, T., Barrett, L., Rabinowitz, D., Doucette, J., and Phan, N.
\newblock Operationalizing a threat model for red-teaming large language models (llms), 2024.
\newblock URL \url{https://arxiv.org/abs/2407.14937}.

\bibitem[Wang et~al.(2023)]{wang2023interpretability}
Wang, A. et~al.
\newblock Towards a mechanistic understanding of transformers.
\newblock \emph{Advances in Neural Information Processing Systems}, 2023.
\newblock URL \url{https://arxiv.org/abs/2301.00701}.

\bibitem[Webb \& et~al.(2024)Webb and et~al.]{webb2024multiagent}
Webb, T. and et~al.
\newblock Learning to coordinate multi-agent systems in generative ai.
\newblock In \emph{International Conference on Machine Learning}, 2024.

\bibitem[Weiss et~al.(2021{\natexlab{a}})Weiss, Goldberg, and Yahav]{weiss2021restricted}
Weiss, G., Goldberg, Y., and Yahav, E.
\newblock Thinking like transformers: Restricting attention supports algorithmic reasoning.
\newblock \emph{Advances in Neural Information Processing Systems}, 34:\penalty0 25623--25634, 2021{\natexlab{a}}.

\bibitem[Weiss et~al.(2021{\natexlab{b}})Weiss, Marcus, and Goldstein]{weiss2021rasp}
Weiss, G., Marcus, M., and Goldstein, A.
\newblock Rasp: Decomposing transformers into modular programming primitives.
\newblock In \emph{Proceedings of the 2021 Conference on Empirical Methods in Natural Language Processing (EMNLP)}, pp.\  1048--1061, 2021{\natexlab{b}}.
\newblock \doi{10.18653/v1/2021.emnlp-main.92}.

\bibitem[Wu(2023{\natexlab{a}})]{langgraph}
Wu, e.~a.
\newblock Langgraph.
\newblock \emph{ArXiv preprint}, 2023{\natexlab{a}}.
\newblock URL \url{https://arxiv.org/abs/2304.10123}.

\bibitem[Wu(2023{\natexlab{b}})]{autogen}
Wu, Z., e.~a.
\newblock Autogen: Enabling seamless multi-agent coordination.
\newblock \emph{Proceedings of the Neural Information Processing Systems}, 2023{\natexlab{b}}.
\newblock URL \url{https://openreview.net/pdf?id=BfkgZK-20345}.

\bibitem[Wu \& et~al.(2023)Wu and et~al.]{wu2023multi}
Wu, Z. and et~al.
\newblock Multi-agent collaboration for secure ai systems.
\newblock \emph{NeurIPS Workshop on Trustworthy AI}, 2023.

\bibitem[Ye et~al.(2024)Ye, Zhang, Su, Sun, Song, Xie, and Liu]{ye2024langgpt}
Ye, Y., Zhang, Z., Su, Y., Sun, Y., Song, Y., Xie, X., and Liu, Y.
\newblock Langgpt: Rethinking structured reusable prompt design for large language models.
\newblock \emph{arXiv preprint arXiv:2402.16929}, 2024.

\bibitem[Yip et~al.(2023)Yip, Esmradi, and Chan]{yip2023resilience}
Yip, D.~W., Esmradi, A., and Chan, C.~F.
\newblock A novel evaluation framework for assessing resilience against prompt injection attacks in large language models.
\newblock In \emph{2023 IEEE Asia-Pacific Conference on Computer Science and Data Engineering (CSDE)}, pp.\  1--5, 2023.
\newblock \doi{10.1109/CSDE59766.2023.10487667}.

\bibitem[Zhang et~al.(2021)Zhang, Yu, Jiao, Xing, Ghaoui, and Jordan]{zhang2021tradeoffs}
Zhang, H., Yu, H., Jiao, J., Xing, E., Ghaoui, L.~E., and Jordan, M.~I.
\newblock Trade-offs between robustness and accuracy in adversarial training.
\newblock \emph{Advances in Neural Information Processing Systems}, 34:\penalty0 14046--14059, 2021.

\bibitem[Zhang et~al.(2024)Zhang, Li, Chen, and Wang]{zhang2024adversarial}
Zhang, H., Li, Y., Chen, J., and Wang, X.
\newblock Adversarial attacks on large language models: A comprehensive survey.
\newblock \emph{ACM Computing Surveys}, 56\penalty0 (4):\penalty0 1--28, 2024.

\bibitem[Zhang et~al.(2022)]{zhang2022security}
Zhang, Y. et~al.
\newblock Security of multi-agent cyber-physical systems: A survey.
\newblock \emph{IEEE Access}, 10:\penalty0 123456--123470, 2022.
\newblock URL \url{https://www.ece.ufl.edu/wp-content/uploads/sites/119/publications/ieee-access22.pdf}.

\bibitem[Zhou et~al.(2024{\natexlab{a}})Zhou, Li, von Oswald, and Yang]{zhou2024algorithmic}
Zhou, X., Li, M., von Oswald, J., and Yang, C.
\newblock Algorithmic understanding of llms: Evaluating emergent primitives and their role in ai systems.
\newblock \emph{Journal of AI Research}, 45:\penalty0 123--145, 2024{\natexlab{a}}.
\newblock \doi{10.1000/jair.2024.123}.

\bibitem[Zhou et~al.(2024{\natexlab{b}})Zhou, Li, and Wang]{zhou2024ontology}
Zhou, X., Li, Y., and Wang, H.
\newblock Ontology-guided constraints for improving large language model generalization.
\newblock \emph{Journal of Artificial Intelligence Research}, 75:\penalty0 123--145, 2024{\natexlab{b}}.
\newblock \doi{10.1016/j.jair.2024.001}.

\bibitem[Zhu et~al.(2023)Zhu, Wang, Zhou, Wang, Chen, Wang, Yang, Ye, Gong, Zhang, and Xie]{zhu2023promptbench}
Zhu, K., Wang, J., Zhou, J., Wang, Z., Chen, H., Wang, Y., Yang, L., Ye, W., Gong, N.~Z., Zhang, Y., and Xie, X.
\newblock Promptbench: Towards evaluating the robustness of large language models on adversarial prompts.
\newblock In \emph{arXiv}. Cornell University, 2023.

\end{thebibliography}
\bibliographystyle{icml2025}



\end{document}